Quantum monitoring the metabolism of individual yeast mutant strain cells when aged, stressed or treated with antioxidant

Aryan Morita[1,2+], Anggrek C. Nusantara[1+], Felipe P Perona Martinez[1], Thamir Hamoh[1], Viraj G. Damle[1], Kiran J. van der Laan[1], Alina Sigaeva[1], Thea Vedelaar[1], Michael Chang[3], Mayeul Chipaux[4*], Romana Schirhagl[1*]

1. Department of Biomedical Engineering, Groningen University, University Medical Center Groningen, Antonius Deusinglaan 1 9713AV Groningen, The Netherlands
2. Department of Dental Biomedical Sciences, Universitas Gadjah Mada, Jalan Denta 1 55281 Yogyakarta, Indonesia
3. European Research Institute for Biology of Aging, University Medical Center Groningen, Antonius Deusinglaan 1 9713 AV Groningen, The Netherlands
4. Institute of Physics, École Polytechnique Fédérale de Lausanne (EPFL), CH-1015 Lausanne, Switzerland

[a]Corresponding author

+ These authors contributed equally

**Abstract**

**Free radicals play a key role in the ageing process. The strongly debated free radical theory of ageing even states that damage caused by free radicals is the main cause of aging on a cellular level. However, free radicals are small, reactive and short lived and thus challenging to measure.**
**We utilize a new technique called diamond magnetometry for this purpose. We make use of nitrogen vacancy centers in nanodiamonds. Via a quantum effect these defects convert a magnetic resonance signal into an optical signal. While this method is increasingly popular for its unprecedented sensitivity in physics, we use this technique here for the first time to measure free radicals in living cells. Our signals are equivalent to T1 signals in conventional MRI but from nanoscale voxels from single cells with sub-cellular resolution. With this powerful tool we are able to follow free radical generation after chemically inducing stress. In addition, we can observe free radical reduction in presence of an antioxidant. We were able to clearly differentiate between mutant strains with altered metabolism. Finally, the excellent stability of our diamond particles allowed us to follow the ageing process and differentiate between young and old cells. We could confirm the expected increase of free radical load in old wild type and *sod1Δ* mutants. We further applied this new technique to investigate *tor1Δ* and *pex19Δ* cells. For these mutants an increased lifespan has been reported but the exact mechanism is unclear. We find a decreased free radical load in, which might offer an explanation for the increased lifespan in these cells.**

Free radicals are involved in many processes in healthy cells and organisms, such as mitochondrial metabolism, cell death and signaling of cells. They are also part of the working mechanism of many drugs. Additionally, they are suspected to play a crucial role in numerous pathogenic conditions including the diseases responsible for most deaths worldwide: cardiovascular diseases, cancer or immune responses to pathogens[1]. Free radicals also play a critical role in ageing. The free radical theory of ageing states that free radicals are the major cause of ageing on the cellular level[2]. Despite their relevance, we know fairly little about them. The reason is that they are produced in low concentrations (ranging from nanomolar to micromolar concentrations) [3,4] and have short lifetimes. Quantifying, identifying and localizing free radicals has been recognized as main bottlenecks to translate free radical biology into biomedical advances[5].

There are several ways to detect free radicals in cells. The different methods can be divided in indirect and direct methods. The indirect ones measure the cellular response towards a certain species. This is done for instance by detecting the expression of certain genes which encode enzymes involved in coping with stress (e.g. superoxide dismutase or catalase)[6]. One way to achieve this goal is via quantitative reverse transcription polymerase chain reaction (RT-qPCR). The major advantage is that these enzymes are specific and one can differentiate between different species. However, one needs to know in advance which enzymes or molecules are involved (which is often unknown). Additionally, the actual radical species are converted into each other[7]. So if you for instance have more catalase, it doesn't necessarily mean that it is the hydrogen peroxide production that went up. Furthermore, it takes some time for a cell to change the gene expression patterns and even more time is needed to adjust proteins expression. Thus, these data always represent the samples history rather than the current situation. Furthermore, spatial and temporal information on where in the cell and when the chemicals were generated is lost.

In contrast imaging is a more direct approach and provides spatially resolved data. Fluorescence imaging allows direct detection of free radicals. Additionally, it is relatively easy and photons can be detected very sensitively (in principle single photon counting is possible). Thus, the spatial resolution that is achieved is only limited by diffraction, down to a few hundreds of nm, which can be even be improved by super-resolution techniques. The assays are based on organic dyes (like 2,7-dichlorofluorescin diacetate (DCFDA) which is used for comparison in this work), which fluoresce when reacting with reactive oxygen species (ROS). However, there are several problems with this approach.

First, although there is some overlap between ROS and free radicals they are not the same (there are some ROS that are not radicals as for instance $H_2O_2$ while there are also some radicals that are not ROS as for instance nitrogen based radicals or radical degradation products of biomolecules). Secondly, the ROS-specific dyes are often consumed in the process and thus distinct time points and no real time curves are obtained. These dyes mostly are not specific to radicals and react with a wide range of different reactive molecules. Only a limited number of specific dyes for radicals are available. There is currently no probe available, which can detect the sum of all radical. Additionally, organic dyes which are used in these techniques usually suffer from photo-bleaching and thus can only be used for short term studies[8,9]. These organic dyes are especially in high concentrations also often somewhat toxic. Another drawback is that organic dyes always measure the history of the sample rather than the current status and the detected value can only increase. So detecting inhibition or the effect of an antioxidant in *real-time* is not possible.

Furthermore, direct techniques also include conventional magnetic resonance imaging (MRI) and electron spin resonance spectroscopy (ESR) that are the gold standard in many different scientific disciplines. These techniques allow for functional contrast. However, especially when high spatial resolution is required or only limited amounts of sample are available, the methods approach their limits. In conventional MRI, approximately $10^{12}$-$10^{18}$ atomic nuclei (or a factor of 1000 less electron spins) are needed to generate an observable signal. This limits the resolution to about 3 $\mu m^3$ at its best[10,11]. In addition, these techniques require large and powerful magnets, which reflect on the price, complexity and availability of the instruments.

Solutions to circumvent these problems are emerging from the field of diamond magnetometry that makes use of defects in diamonds which change their optical properties based on their magnetic surroundings[12,13,14].

The most prominent of these defects, the so-called nitrogen vacancy center (NV center), is so sensitive that even the small magnetic signal of single electrons[15] or nuclear spins[16,17] can be detected. In physics, this technique has successfully been used for a number of applications including magnetic characterization of materials under high hydrostatic pressures[18,19,20], 2 dimentional materials magnetic nanoscale microscopy[21] or high speed study of magnetic domain walls[22], magnetic fields of nanoparticles[23] or the presence of molecules on the diamond surface[24,16,25]. Diamond magnetometry has also been used to measure spin labels in solution or even in a biological environment[26]. A study from Ermakova et al (2013)[27] showed the detection of iron in ferritin (a protein family found in many living organisms). Steinert et al. were able to record data from cells that were fixed, sliced and labeled with contrast agents[28]. Le Sage et al. recorded the magnetic signature of magnetosomes (naturally occurring iron nanoparticles) in living magnetotactic bacteria[23]. Glenn et al. investigated cells that were labeled with magnetical nanoparticles[29]. This way they were able to quantify biomarker expression at a single cell level.

Here we use diamond magnetometry for the first time for measuring the natural metabolism in living yeast cells. Such yeast cells are popular in ageing research for several reasons. Many processes in ageing and coping with stress are highly conserved between different organisms including yeast. Unlike in most other cells is possible to differentiate between young and old cells by size. Additionally, yeast can age in two different ways: replicatively (by dividing) and chronologically (by spending time in a non-replicative stage). Both are used extensively to model ageing in different types of human cells.

In our article, we detect the free radical formation during stress responses or the addition of antioxidants and chronological ageing of yeast cells. Despite the relevance, such information remained unavailable so far.

**Results and Discussions**

In this study, we provide a new approach for detecting free radicals during stress conditions and the aging process in yeast. Figure 1 shows a typical workflow for a magnetometry experiment. After uptake into yeast cells as described earlier[30], we first collect confocal images of yeast cells. By imaging z-stacks (Figure 1 (A)) we determined if a particle was indeed inside the cell. Once a diamond nanoparticle was identified, we performed a magnetometry measurement. More specifically we used a so-called relaxation or T1 measurement of the NV center spins. This technique, which requires optical manipulation only, is sensitive to the spin noise generated by free radicals[31]. The pulsing sequence that is needed is shown in Fig 1 (B) and a typical measurement in Fig 1 (C). During this sequence the NV centers are brought into the (bright, polarized) ground state. After different amounts of time (=dark time τ) we then probed whether or not the NV centers remained in this state. If there is spin noise (in our case from free radicals) present, this process occurs faster. The characteristic time this process takes is called relaxation time or T1 time. It gives a measure for the concentration of radicals in the surrounding of the diamond particle. After an initial measurement of the T1 of a nanodiamond in a cell, we perform a biological intervention on the cell and observe the effect on the T1 (on the same nanodiamond, in the same cell). Each set of experiment was reproduced 4 or 5 times.

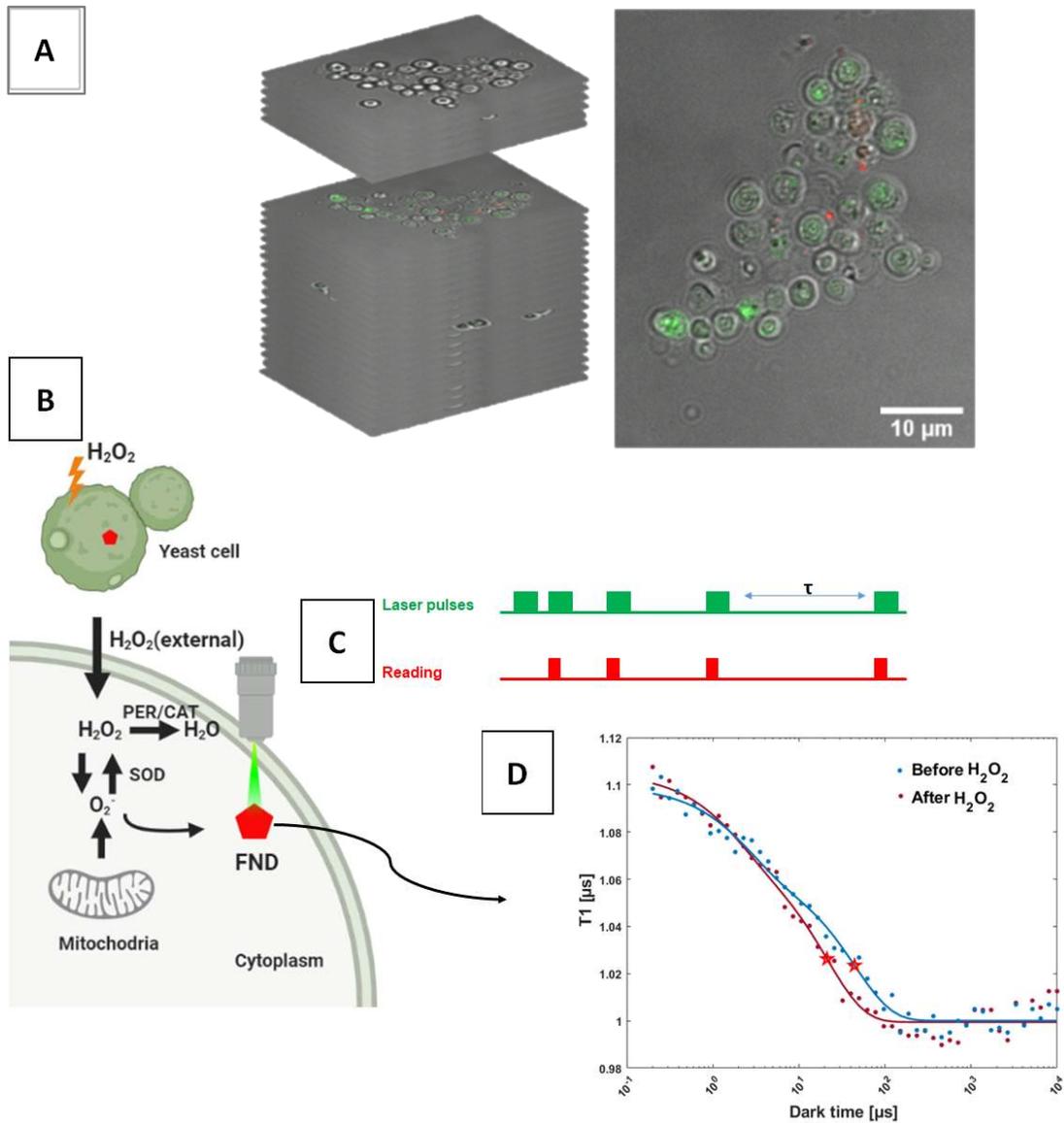

**Figure 1**. Concept of detecting radical formation in yeast. (A) shows a z-stack of yeast cells with FNDs inside (red: FNDs, green: green fluorescent protein (GFP) (B) shows a schematic of a yeast cell triggered by $H_2O_2$ to induce the oxidative stress response by inducing free radical generation. FNDs with NV center sense the magnetic noise that is produced by free radicals. (C) shows the pulsing sequence that is produced in a home-built diamond magnetometry setup. We use laser pulses at a varying distance to pump NV defect in the ground state. After different dark times we record the photoluminescence signals to determine how long the NV center remains in this polarized state. In presence of noise from radicals surrounding the NV centers this process occurs faster. Thus, this T1 time gives a measure for the concentration of radicals in the surrounding. (D).

**Comparing wild-type and knock out strains before and after ageing**

We used 4 different yeast strains and observed the free radical load during the aging process. We chose wild type cells and three knockout mutants *sod1Δ*, *tor1Δ,* and *pex19Δ*, which are expected to have an impaired metabolism. Superoxide dismutase (SOD) is an enzyme that is found in mitochondria (MnSOD (encoded by the *SOD2* gene)) and also in the cytoplasm (Cu/ZnSOD (encoded by *SOD1* gene))[32]. SOD controls the level of $H_2O_2$ and engages in metabolic regulation as a response to the presence of oxygen and glucose[33,34]. SOD also binds to two casein kinases through localized production of $H_2O_2$. In short, SOD is very crucial for dismutating $O_2^-$ free radicals[35]. Deletion of the *SOD1* gene in yeast (*sod1Δ*) causes extensive damage of DNA and oxidative stress in cells[36]. This means free radical production cannot be inhibited and the cells cannot degrade free radicals. As consequence, intracellular free radical levels are expected to increase.

The *tor1Δ* mutant was chosen because *TOR1* (target of rapamycin) regulates cell growth and metabolism. It has been found in several cellular locations but mainly on the vacuolar membrane[37,38]. It also regulates the transcription factors Msn2 and Msn4, which are responsible for different stress responses[39,40]. Deletion of *TOR1* in yeast (*tor1Δ*) decreases cell metabolism.

The *PEX19* (peroxin 19) is gene that is responsible for peroxisome membrane biogenesis[41]. Such a defect on peroxisome membrane biogenesis will affect the metabolism of peroxisomes which are responsible for the generation of superoxide and nitric oxide[42]. However, for both TOR1 and PEX19 it is unknown how this exactly influences the free radical load. The differences in T1 values of the 4 yeast strains can be found in Figure 2.

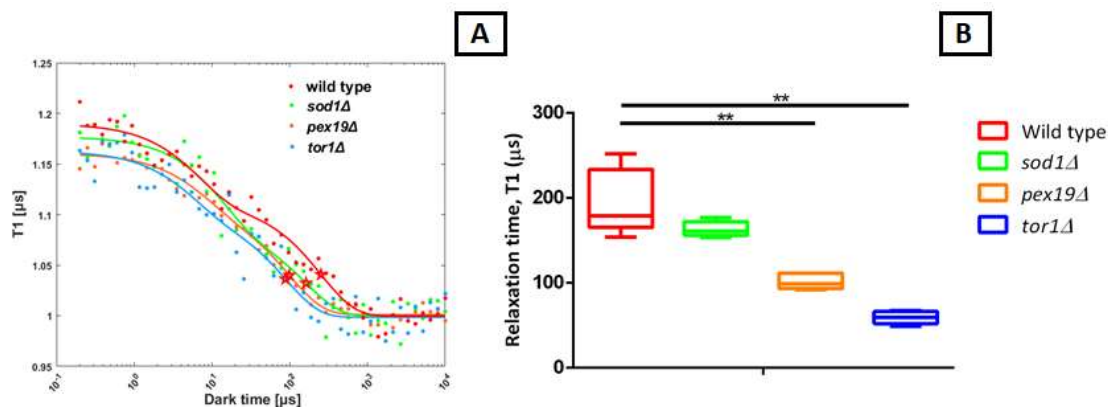

**Figure 2.** Representative T1 curves in cells before any interventions **(A)** and average values **(B)** in 4 different yeast strains. Wild type strain (BY4741) (red) has the highest T1 signal (meaning the lowest free radical load) compared with the 3 isogenic mutants (*sod1Δ* (green), *pex19Δ* (orange), and *tor1Δ* (blue)). All measurements were performed at room temperature in living yeast cells. The data were analyzed by using one way ANOVA followed by Tukey post hoc test. Analyzes were done against the wild type group. Significance level was set to 0.05. ** indicates P≤0.01.

All mutant strains have lower T1 values and thus higher free radical loads compared to the wild type strain. The lowest value was detected in *tor1Δ*. We also compared all our measurements with 3 conventional methods, the MTT assay (a measure for metabolic activity), the DCFDA assay (measuring ROS production) and the HPF assay (measuring hydroxyl radicals (*OH)). In Figure S1 of the supplementary information we show the results for the standard techniques. While all other trends are the same we find that the metabolic activity in *tor1Δ* was lowest. While comparing ROS production level, *sod1Δ* showed the highest ROS production compared to others. Looking at [*OH] level between those 4 strains, *sod1Δ* also showed the higher level of [*OH] and *tor1Δ* has the lowest [*OH]. From the data in Figure S1 (and even more so in the detailed comparison of the methods in Figure S11) it is clear that the data that T1 provides is unique.

While the trends are often correlated, none of them measures the exact same as relaxometry (the local concentration of all radicals at a given moment). The DCFDA and MTT measurements are dominated by the non-radical species, which are way more abundant. HPF on the other end only measures a small subset of the radicals that are created.

Additionally, the existing methods do not allow following the same cell during the entire experiment. Thus, these results are always an average of a population of cells whereas T1 signals are from one specific cell. As mentioned earlier there is also a fundamental difference in what kind of data one obtains. The existing methods always reveal the history of the sample and capture everything that has been generated between adding the compound and the measurement. Since these compounds can diffuse, the location (especially for long measurements) is very inaccurate. Our relaxometry based approach on the other hand measures the current situation.

**Investigating stress responses and ageing**

Next, we observed the yeast's free radical response to the addition of a chemical stress. To this end 1% $H_2O_2$ was used because it is highly diffusible and below a lethal concentration for yeast.

Additionally, it is commonly used as a bleaching agent and also to induce stress in cells[43,44]. We first measured the initial condition of the cells (before any intervention). We then triggered them with $H_2O_2$ and measure in the same cells at the same location again. As expected, the 4 strains responded differently when they were exposed to stress (see Fig. 3). In Figure 3 A, B, C, and D we display the evolution of T1 caused by $H_2O_2$. We also compared these results with metabolic activity, ROS production and *OH radical production (supplementary information figure 2,3, 4 and 12).

We next investigated the effect of 2 stress factors: ageing and $H_2O_2$ as chemical stress. Here, we compare between T1 before and after $H_2O_2$ in aged cells (see Figure 3 **(E), (F), (G), (H)**). It is worth mentioning that the high photostability of diamonds allows for a very powerful control, as the reference measurements can be performed on the same particle in the same location within the same cell. In contrast, with organic dyes/other nanoparticles, one usually has to use the data from different cells as a baseline for the experiments.

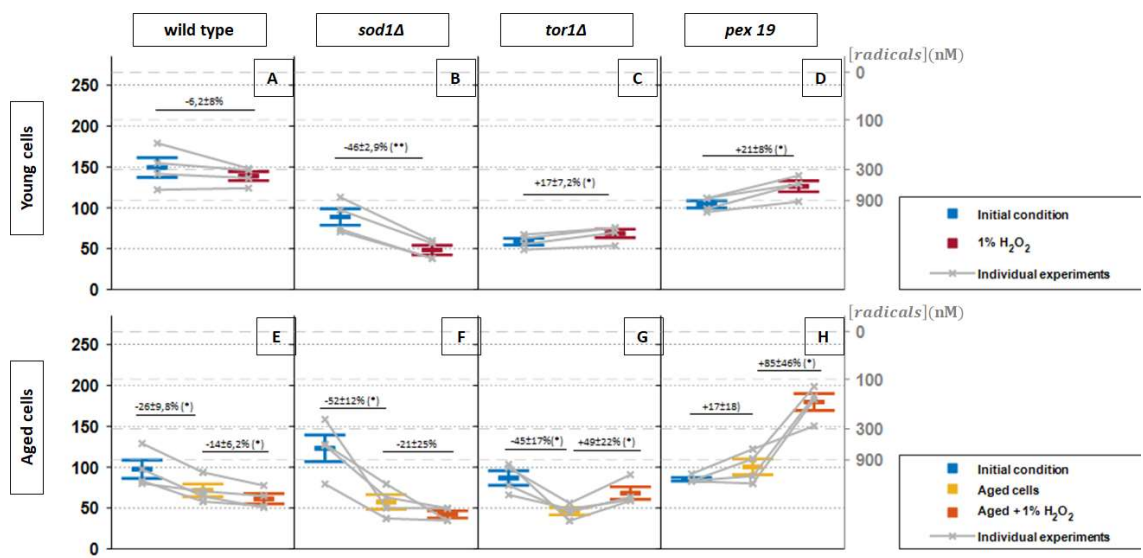

**Figure 3**. Aging and stressed cells. We plot the T1 response (left black axis) of internalized nanodiamond in 4 yeast strains. As an indication, the left grey axis represents the approximated concentration determined from a calibration with *OH radicals in solution from previous work[45]. For young cells we compare the T1 in the cell before (blue) and after adding 1% $H_2O_2$ to promote oxidative stress (red) in **(A), (B), (C)** and **(D).** Then we compare young cells (blue) with themselves after 24 h of ageing (yellow) in **(E), (F), (G), (H)**. In the same figures we also show the effect of 1% $H_2O_2$ in addition to ageing. The error bars represent standard errors. The experiments were repeated 4 times. The grey lines follow individual experiments. Data were analyzed by using paired T-test against the previous group. Significance level was set at 0.05 (*P≤0.05; **P≤0.01)

In young cells, both wild-type and s*od1Δ* have lower T1 after adding $H_2O_2$ while *tor1Δ* and *pex19Δ* show slightly higher T1 values. This can be seen from the average values but also from the grey data points generated each from a single cell experiment. We can explain these findings in the following way. *Sod1Δ* cells already contain a high amount of free radicals as a response to deletion of *SOD1*. When 1% $H_2O_2$ was added to the cells, it triggered mitochondria to produce more radicals. This can be concluded from the decreasing of T1 values. On the contrary, *tor1Δ* and *pex19Δ* behave differently than the wild type and *sod1Δ* strains. A previous study showed that *TOR1* inhibits transcription of stress-responsive elements and controls a starvation response[39]. Semchyshyn and Valishkevych (2016)[46] showed that deletion of the *TOR1* gene activates glyoxalase 1 (*GLO1*), which is associated with the metabolism of carbohydrates and $H_2O_2$. As a result, $H_2O_2$ is metabolized and yeast cells become more resistant to oxidative stress. Mutation of the PEX19 gene in yeasts means a lack of peroxisomes[47]. Peroxisomes are cell organelles which function as pro and antioxidants. According to Pascual-Ahuir and coworkers (2017)[48], peroxisomes are producing oxidase and catalase to combat oxidative stress. In this study, *pex19Δ* revealed that there is lower radical production after adding 1% $H_2O_2$. This means that the mutation of the PEX19 gene increases the resistance of yeasts to oxidative stress.

We then focused on chronological aging where the cells entered the stationary phase of growth and are deprived of nutrients. Accumulation of oxidatively damaged proteins and ROS has been observed during chronological ageing[49,50]. To increase the longevity of the cells and stress response during nutrition scarce conditions, several pathways (Sch9, TOR, and Ras/cAMP-dependent PKA) are down-regulated[51]. In our experiments, wild type, *sod1Δ*, and *tor1Δ* have T1 value lowered with aging, which indicates a higher free radical load. According to the free radical theory of aging[52] these results are expected. Compared to wild type and *sod1Δ*, *tor1Δ*, which has an increased chronological lifespan[53], shows a slightly lower T1 value after aging, and when aging cells were stressed by $H_2O_2$, T1 values slightly increase. Comparing T1 values with metabolic activity (supplementary Figure S2), both *sod1Δ* and *tor1Δ* in young and aged stages decreased their metabolic activity profile in the presence of 1% $H_2O_2$. Interestingly, aged *pex19Δ* has even a slightly lower radical load compared to their initial young condition. This finding provides an explanation for the increased chronological lifespan which has been found for these cells[54]. We attribute this finding to the elevation of hydrogen peroxide, which activates superoxide dismutase (SOD) as scavenger of superoxide anions. A detailed comparison with the existing methods is shown in Figure S12 of the supplementary.

**Effect of adding antioxidant on free radical loads**

In yeasts, there are enzymatic and non-enzymatic defense systems against chemical stress. In enzymatic systems, several enzymes (e.g. SOD and catalase) are involved in removing oxygen radicals and repair the damage from oxidative stress[32]. Non-enzymatic systems involve glutathione (GSH) and endogenous antioxidants[55]. The non-enzymatic system acts as radical scavenger, which maintains the cellular redox state. Although yeast cells produce natural radical scavengers, adding external anti-oxidant supplements helps them to survive oxidative stress. Here, we investigate how the radical production changes after antioxidant treatment. Figure 4 shows the effect of the common anti-oxidant L-ascorbic acid.

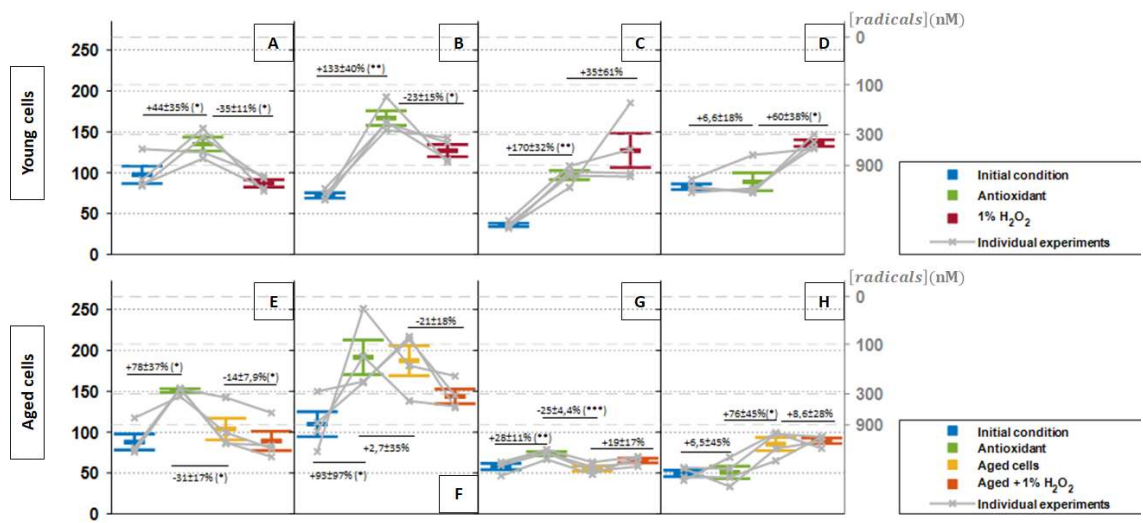

**Figure 4**. Free radical response to ageing in presence or absence of an antioxidant/stressor. In figure (A), (B), (C), and (D) we first compare T1 values (left black axis). The left grey axis represents the approximated concentration determined from a calibration with *OH radicals in solution from previous work[45]. We investigated wild type, $sod1\Delta$, $tor1\Delta$, and $pex19\Delta$ cells before (blue) and after antioxidant treatment (green). We also show the response after subsequent addition of 1% $H_2O_2$. In (E), (F), (G) and (H), we additionally display the evolution of T1 values after ageing. We compare the initial condition with young cells to antioxidant treated cells followed by ageing and adding 1% $H_2O_2$. The error bars represent the standard error, above and below the average. The experiments were repeated

4 times. Data were analyzed by using paired T-test against the previous group. Significance level was set at 0.05 (*P≤0.05; **P≤0.01; ***P≤0.001)

In young cells, adding anti-oxidant increases the T1 values (i.e. reduces the free radical load in cells). In the wild-type **(A)** and *sod1Δ* **(B)**, adding anti-oxidant before adding 1% $H_2O_2$ does not inhibit radical production in cells. This is indicated by decreasing of T1 values after 1% $H_2O_2$ has been added. On the other hand, adding anti-oxidant in *tor1Δ* **(C)** and *pex19Δ* **(D)** reduces free radicals formation. In aged cells, anti-oxidant was added before the cells were allowed to age. In all cases, the aged cells that were treated with the antioxidant are better off after the addition of $H_2O_2$, as compared to the results in Fig 3. Since all the strains have been grown in glucose supplemented medium, they are not able to produce L-ascorbic acid[56] but they can accumulate external ascorbate through a hexose uptake system[57]. Our findings thus offer an explanation for the findings of Saffi and coworkers (2006)[58] who reported that adding L-ascorbic acid supplement in wild type and *SOD1* deficient strains before oxidative stress conditions increases survival. L-ascorbic acid can restore the cellular activity of yeast due to the scavenging of superoxide radicals. In supplementary Figure S5 and S6, adding anti-oxidant increases metabolic activity and decreases ROS concentration in all strains compared to the situation without anti-oxidant groups. However, when the cells are aged, anti-oxidant supplement has no significant effect for improving metabolic activity levels. Anti-oxidant supplement also only slightly reduces [*OH] in all strains both in young and age stage (Figure S7 and for a more detailed comparison of methods see Figure S13)

**Conclusions**

Free radicals are central in oxidative stress responses and the ageing process. Since these molecules are highly reactive and have short life times, there are limited techniques that can be used to measure the free radical load. Our study offers a new and unique technique, which can measure the productions of free radical by single cells in real time with high spatial resolution. With the T1 signal, we were able to clearly differentiate the response of mutant strains, young and aged cells, in the presence or absence of $H_2O_2$ and anti-oxidants.

Using NV centers in FNDs as a tool to convert magnetic noise from free radicals into optical signals provides a better understanding about stress responses of the cells during oxidative stress conditions and the ageing process. Using this new tool we are able to provide an explanation for increased longevity that has been reported for *tor1Δ* and *pex19Δ*.

We believe that this work paves the way to faster and more efficient free radicals studies in living cells, reducing essay size down to the single cells level. This new information is not just fundamentally interesting but also opens up opportunities to evaluate the effect of mutation as well as new drugs and their working mechanisms.

**Methods**

**Diamond starting material.** All experiments in this article were done using commercially available fluorescent nanodiamonds (FNDs) with a hydrodynamic diameter of 70 nm from Adamas Nanotechnology (USA), which seem to be a good compromise. There are two factors that are important to balance here. The larger the particle, the more NV centers can contribute to the measurement. Since every measurement is already an average of many centers at different locations within the particle, the particles are very similar to each other and thus the smaller the error bars are. However, if the particle becomes too large, the NV centers in the center of the particle do not feel the signal of the radicals any more due to their large distance from the surface. These particles are produced by grinding HPHT diamonds and irradiation with 3MeV electrons at a fluence of $5*10^{19}$ e/cm$^2$. As a result these particles contain an average of 500 NV centers per diamond (determined by EPR by the manufacturer)[59]. The surface chemistry of the FNDs is oxygen-terminated, as a result of an acid treatment by the manufacturer. These particles have already been characterized extensively in previous literature[60,61].

**Diamond uptake in yeast cells.** Budding yeast *Saccharomyces cerevisiae* cells (of the BY4741 strain background)[62] were grown in synthetic dextrose (SD) medium (Formedium, UK) until they reached midlog phase (optical density 600 nm (OD$_{600}$) approximately 1,05). Since yeast cells are surrounded by a rigid cell wall and do not naturally take up diamond particles, it is necessary to remove or permeabilize the cell wall[30]. Removing the cell wall leads to a so-called spheroplast.
First, the cells were washed with sterile demineralized water for 5 minutes at 2500 xg at 10°C. After the supernatant was discarded, 20 ml of sterile 1 M D-sorbitol were added to the cells followed by centrifugation for 5 minutes at 2500 xg at 10°C. Sorbitol is a reduced medium for yeast cells, which does not cause diamond aggregation[63]. Then the supernatant was discarded. The cell pellet was diluted in 20 ml of sterile SPEM buffer (consisting of 1 M D-sorbitol, 10 mM EDTA, and 10 mM sodium phosphate) and 40 μl of zymolyase 20T (Amsbio, UK). Then 30 μl of β-mercaptoethanol (Sigma, Netherlands) were added. The mixture was incubated at 30°C for 30 minutes while shaking at 75 rpm. The spheroplasting process was monitored with a light microscope (native yeast cells are darker than spheroplast cells). The process was stopped by adding 20 ml of sterile 1 M D-sorbitol followed by centrifugation at 1000 xg for 5 minutes at 10°C. The pellet then was mixed with 2 ml of sterile STC buffer (consisting of 1 M D-sorbitol, 10 mM Tris HCl, 10 mM CaCl$_2$, and 2,5 mM MgCl$_2$). The mixture was incubated for 20 minutes at room temperature[64]. Finally, 50 μl of 4 μg ml$^{-1}$ FNDs in 1 M D-sorbitol was added to the suspension followed by a 10 minute incubation at room temperature.
The isogenic mutant strains *sod1Δ*, *tor1Δ,* and *pex19Δ* were obtained from the Yeast Knockout Collection[65]. They were maintained on solid yeast extract-peptone-dextrose (YPD) supplemented with 2% D-glucose. For all the experiments, the cells were grown in SD medium supplemented with 2% D-glucose followed by spheroplasting for the diamond uptake protocol.

**Sample preparation for diamond magnetometry.** Yeast cells with internalized FNDs were seeded in concanavalin-A coated glass-bottom dishes. The function of concanavalin A is to promote cell adhesion to the glass surface and thus limiting cell movement. For coating, the glass surface of the dishes (Grainer) were submerged in 0,1 mg ml$^{-1}$ concanavalin-A (Sigma, The Netherlands) in sterile demineralized water. The dishes were incubated at 37°C overnight in static condition. On the following day, the dish was washed using sterile demineralized water followed by drying in a 37°C incubator[66]. The first step of the experiment was finding the FNDs and confirming their location inside the cell using Z-stack imaging.

**Diamond magnetometry.** To perform diamond magnetometry we utilized fluorescent defects in diamond called NV centers. Using this technique, the magnetic noise of the surrounding medium is read by optical means. These experiments were done using a home-made magnetometry setup. The setup, which has been described before[66] is in principle a confocal microscope with a few changes which are described below. For light collection we used a 100x magnification oil objective (Olympus, UPLSAPO 100XO).

We implemented the ability to pulse the laser with an acousto-optical modulator (Gooch & Housego, model 3350-199) to conduct the pulsing sequence that is shown in Figure 1.

There are several modes that can be used in diamond magnetometry. These modes (including the rotating frame magnetometry[67], T2[13], CPMG[68]) require different pulsing sequences. These sequences require an alignment of an external magnetic field with the NV centers. This would be problematic for a rotating particle. Here we chose the relaxometry or T1 sequence. Probably the biggest advantage of this sequence is that it does not require microwaves. In biological samples the water in the medium absorbs the microwaves. This not just deteriorates the signal but also causes heating. During a T1 measurement the NV center is pumped with a laser pulse into the (bright) ground state (see Figure 1 (C)). Then we probe after different dark times, if the NV centers are still in this ground state. The time it takes to reach the equilibrium condition (relaxation time) is linked to the presence of the concentration of free radicals. To estimate which concentration of radicals causes a certain T1 time we used a calibration where *OH radicals were created in a controlled way[45].

To perform a T1 measurement, a train of 5 µs green laser pulses (532 nm) with dark times ($\tau$) between 200 ns to 10 ms was used to excite the NV centers. The detection is done using an avalanche photodiode (APD) (Excelitas, SPCM-AQRH) after passing through a 550 nm long-pass filter. To obtain a sufficient signal to noise ratio we repeated the pule sequence 10000 times for each T1 measurement. Each T1 measurement took around 16 minutes. The optically detected T1 measurements are equivalent to T1 signals in conventional magnetic resonance imaging (MRI) but for nanoscale voxels. There are several considerations that determine the choice of laser power. The laser power has to be low enough to not damage the cells. On the other hand the power needs to be high enough to polarize the NV centers sufficiently. A good compromise was found at 50 µWatt measured on top of the objective lens. A common strategy in the field is using diamonds, which contain single defects[69,31]. This has several advantages. The particles can be small and single defects can be very precisely manipulated. However, this approach usually requires spending significant amounts of time on searching for a superb defect. These highly preselected defects have fantastic sensing properties. Once the perfect defect is found it is typically reused for a long time. Unfortunately, this approach doesn't work in a biological system. We need to use whichever particle we can get in the cells and if we would preselect from these, we could only use the particle once and discard it with the biological samples, which need to be disposed after each experiment. Even worse is the large spread one can observe in defect quality for single defects. This would make it impossible to compare different experiments on different cells and different particles. To circumvent this problem we used nanodiamonds with ensembles. These are easier to find on top of the background fluorescence. Since every particle contains around 500 NV centers we always measure the sum of many different NV centers, which are more or less close to the surface and to each other. The result is, that the measurements with different particles are way more reproducible than with single centers at the expense of relaxation time.

Since diamond particles move inside the cells it is crucial to use a tracking algorithm. The purpose of the tracking algorithm is to find the new position of the particle and move the laser there. The procedure has to be repeated more often if the particle is moving faster. For the conditions in this article we found that it is necessary to track the position of FND every 5 seconds during a T1 measurement. If the particle is lost this results in a sudden drop in fluorescent counts.

**Magnetometry data analysis.**

The model that was used to fit the T1 data is described in equation (1):

$$PL(\tau) = I_{inf} + C_a\, e^{-\tau/T_a} + C_b\, e^{-\tau/T_b} \qquad (1)$$

T1 = max(Ta, Tb)

This model is a bit different than the fitting model for single NV[31] because we are using FNDs containing ensembles of NV centers. The relaxation time of an ensemble is approximated from two component which are the NV centers that are closer to the surface (Ta) and the NV centers that are deeper in the crystal (Tb)[70]. Between these two constants, the longer T1 time was selected for quantification because it is more sensitive to changes of the NVs' surrounding. Data were processed with MatLab software version R2018b.

**Cell aging behavior.** The behavior of aging cells was monitored in this study. After spheroplasting, cells were washed with sterile phosphate buffered saline (PBS) pH 7,4 followed by centrifugation at 1000 xg at 10°C for 5 minutes. Cells were then kept in sterile water for 24 hours at 30°C in an incubator. At the desired time, cells were washed and tested with magnetometry or conventional methods (MTT, $H_2$DCFDA, and HPF).

**Effect of adding antioxidant.** An antioxidant was used for preventing or reducing radical formation during stress conditions and during the aging process in yeast cells. In this study, 0,025 mM of L-ascorbic acid (Sigma, Netherlands) has been used as antioxidant. The radical load was determined by magnetometry while conventional methods (MTT, $H_2$DCFDA, and HPF assays) were used for comparison.

**Comparison with established techniques.** To compare magnetometry data with existing methods several assays were performed which are described in the following.

**Morphological changes after stress condition**. Morphological changes on the surface of yeast spheroplast cells that have been exposed with 1% $H_2O_2$ compared to cells without any treatment were visualized by using a bioscope catalyst atomic force microscope (AFM) (Bruker, USA).
To immobilize yeast spheroplasts on a hydrophilic glass slide, 1x1 cm glass slides were sterilized by using 70% ethanol[71]. Then 0,1 mg concanavalin- A (Sigma, Netherlands) per mL sterile water was spread on the surface of the glass slides. After the drying process, they were placed in sterile 12-well plates (Grainer). The cells were incubated at 37°C overnight and washed with sterile demineralized water. Then yeast spheroplast cells were seeded on the concanavalin-A coated glass slides and incubated at 30°C for 3 hours. After that, cells were exposed to 1% $H_2O_2$ and incubated for 1 hour at 30°C. Glass slides with cells on top were taken from the well plates and dried at room temperature for 3 hours before observing by AFM. The results are shown in the supplementary (Fig. S10)

**Metabolic activity measurement.** Metabolic activity was observed by using MTT [3-(4,5-dimethylthiazol-2-yl)-2,5 diphenylterazolium bromide] (Sigma, Netherlands). The assay was modified from the protocol established by Teparić et al (2004)[72]. To perform the assay, spheroplast cells were washed with sterile PBS pH 7,4. Cells were divided into 3 groups and placed in a sterile tube with a volume of 1 ml each tube. Cells were mixed with 1% hydrogen peroxide ($H_2O_2$) (Merck, Netherlands) and 4µg ml$^{-1}$ FNDs. In the control group only sterile demineralized water was added. Then 100 µl of 5 mg ml$^{-1}$ MTT solution in sterile PBS was added to the cells from the different groups. They were wrapped with tin foil and incubated for 1 hour at 30°C. After that, cells were resuspended with 1 ml 2-

propanol (Merck, The Netherlands), agitated for 10 minutes, and centrifuged at 1000 xg for 5 minutes at 10°C. 1 ml of supernatant was taken from all groups and placed in sterile 96-well plates (Grainer). Finally, the plates were measured with a microplate reader (Fluostar optima, Germany) at 540 nm. All the measurements were performed in triplicates.

**Measuring reactive oxygen species (ROS) with fluorescence based probes.** ROS in yeast spheroplasts were measured by using a $H_2DCFDA$ kit (Thermofisher, Netherlands). The cells were placed into 1,5 ml sterile tubes and centrifuged at 1000 xg for 5 minutes. The supernatant was discarded and the cells were washed with sterile PBS pH 7,4 followed by centrifugation at 1000 xg for 5 minutes. After discarding the supernatant, cells were diluted with sterile PBS pH 7,4 and transferred to sterile 96- well plates. The $H_2DCFDA$ with a concentration of 10µg ml$^{-1}$ was added in each well and the cells were incubated for 2 hours at 30°C (the plates were wrapped with aluminium foil to avoid the light exposure). The cells were treated with 1% $H_2O_2$; 4 µg ml$^{-1}$ FNDs; and cells without treatment were used as control. Finally, the fluorescence intensity was measured by a microplate reader (Fluostar optima, Germany) at 540 nm as describe by Gallardo and co-workers (2013)[73]. All the measurements were performed in triplicates.

**Measuring hydroxyl radical (*OH) with fluorescence based probes.** Hydroxyl radical generation was measured by HPF (hydroxyphenyl fluorescein) (Thermofisher, Netherlands). Yeast cells were put in 1,5 ml sterile tubes then centrifuged at 1000 xg for 5 minutes at 10°C. After discarding the supernatant, cells were washed with sterile PBS pH 7,4 followed by centrifugation at 1000 xg for 5 minutes. 10 µM HPF probes in sterile PBS were mixed with the cells and incubated for 1 hour at 30 °C (all tubes were wrapped with aluminum foil to avoid light exposure). Cells were centrifuged at 1000 xg for 5 minutes to remove free HPF followed by diluting the cells into sterile PBS. After placing the cells into 96-well plates (100 µl/well), they were treated with 1% H2O2; 4 µg ml$^{-1}$ FNDs; and cells without treatment were used as control. Fluorescence intensity was measured by a microplate reader (Fluostar optima, Germany) at 495/520 nm. All the measurements were performed in triplicates.

**Statistical analysis.** Continuous data are presented as mean ± standard errors and analyzed by paired T-Test against the previous group. For MTT, $H_2DCFDA$, and HPF without anti-oxidant data were analyzed by using one-way ANOVA followed by Tukey post hoc test and anti-oxidant effect for all measurements were analyzed by using two-way ANOVA followed by Sidak post hoc test. Statistical analysis was performed using GraphPad Prism version 8 software (GraphPad Inc.). P value <0.05 was considered to indicate significance statistic value.

**Acknowledgment**

**A.M** acknowledges a Lembaga Pengelola Dana Pendidikan (LPDP) scholarship from the Republic of Indonesia. **A.C** thanks Graduate School of Medical Sciences University of Groningen for a PhD scholarship. This research was funded by FOM projectruimte, grant number FOM-G-36. **R.S**. is thankful for support via the ERC starting grant ERC-2016-StG-714289.

**Author contributions**

**A.M** and **A.C** contributed equally. **A.M** and **A.C** designed and performed experiments also analyzed the data. **F.P** built the complete setup. **T.H** contributed in preliminary studies, **V.D** performed AFM measurements. **A.S** designed the figures, **T.V** analyzed relaxometry data, **K.L** involved in yeast cells experiment, **M.Cha** provided yeast strains, **M.Chi** was involved in designing the measurement method, building the equipment and analyzing the data and **R.S** planned the experiments and leads the research group. All authors were involved in writing and reviewing the manuscript.

1. Nathan, C. & Cunningham-Bussel, A. Beyond oxidative stress: An immunologist's guide to reactive oxygen species. *Nat. Rev. Immunol.* **13**, 349–361 (2013).
2. Harman, D. The Free Radical Theory of Aging. *Antioxid. Redox Signal.* **5**, 557–561 (2003).
3. Halliwell, B. Free radicals, antioxidants, and human disease: curiosity, cause, or consequence? *Lancet* **344**, 721–724 (1994).
4. Paardekooper, L. M. *et al.* Human monocyte-derived dendritic cells produce millimolar concentrations of ROS in phagosomes per second. *Front. Immunol.* **10**, 1–9 (2019).
5. Phaniendra, A., Jestadi, D. B. & Periyasamy, L. Free Radicals: Properties, Sources, Targets, and Their Implication in Various Diseases. *Indian J. Clin. Biochem.* **30**, 11–26 (2015).
6. Diguiseppi, J. & Fridovich, I. The Toxicology of Molecular Oxygen. *Crit. Rev. Toxicol.* **12**, 315–342 (1984).
7. Nimse, S. B. & Pal, D. Free radicals, natural antioxidants, and their reaction mechanisms. *RSC Adv.* **5**, 27986–28006 (2015).
8. Mason, R. P. Imaging free radicals in organelles, cells, tissue, and in vivo with immuno-spin trapping. *Redox Biol.* **8**, 422–429 (2016).
9. Wardman, P. Fluorescent and luminescent probes for measurement of oxidative and nitrosative species in cells and tissues: Progress, pitfalls, and prospects. *Free Radic. Biol. Med.* **43**, 995–1022 (2007).
10. Ciobanu, L., Seeber, D. A. & Pennington, C. H. 3D MR microscopy with resolution 3.7 μm by 3.3 μm by 3.3 μm. *J. Magn. Reson.* **158**, 178–182 (2002).
11. Glover, P. & Mansfield, P. Limits to magnetic resonance microscopy. *Rep. Prog. Phys.* **65**, 1489–1511 (2002).
12. Gali, A., Fyta, M. & Kaxiras, E. Ab initio supercell calculations on nitrogen-vacancy center in diamond: Electronic structure and hyperfine tensors. *Phys. Rev. B - Condens. Matter Mater. Phys.* **77**, 1–28 (2008).
13. Schirhagl, R., Chang, K., Loretz, M. & Degen, C. L. Nitrogen-Vacancy Centers in Diamond: Nanoscale Sensors for Physics and Biology. *Annu. Rev. Phys. Chem.* **65**, 83–105 (2014).
14. Chipaux, M. *et al.* Nanodiamonds and Their Applications in Cells. *Small* **1704263**, 1–25 (2018).
15. Grinolds, M. S. *et al.* Nanoscale magnetic imaging of a single electron spin under ambient conditions. *Nat. Phys.* **9**, 215–219 (2013).
16. Mamin, H. J. *et al.* Nanoscale nuclear magnetic resonance with a nitrogen-vacancy spin sensor. *Science (80-. ).* **339**, 557–560 (2013).

17. Cujia, K. S. *et al.* Tracking the precession of single nuclear spins by weak measurements. *Nature* **571**, 230–233 (2019).
18. Hsieh, S. *et al.* Imaging stress and magnetism at high pressures using a nanoscale quantum sensor. *Science (80-. ).* **366**, 1349–1354 (2019).
19. Yip, K. Y. *et al.* Measuring magnetic field texture in correlated electron systems under extreme conditions. *Science (80-. ).* **366**, 1355–1359 (2019).
20. Lesik, M. *et al.* Magnetic measurements on micrometer-sized samples under high pressure using designed NV centers. *Science (80-. ).* **366**, 1359–1362 (2019).
21. Thiel, L. *et al.* Probing magnetism in 2D materials at the nanoscale with single-spin microscopy. *Science (80-. ).* **364**, 973–976 (2019).
22. Juraschek, D. M. *et al.* Dynamical Magnetic Field Accompanying the Motion of Ferroelectric Domain Walls. *Phys. Rev. Lett.* **123**, 127601 (2019).
23. Le Sage, D. *et al.* Optical magnetic imaging of living cells. *Nature* **496**, 486–489 (2013).
24. Loretz, M. *et al*. Nanoscale nuclear magnetic resonance with a 1.9-nm-deep nitrogen-vacancy sensor. *Appl. Phys. Lett.* **104**, 3–8 (2014).
25. Staudacher, T. *et al.* Nuclear magnetic resonance spectroscopy on a (5-nanometer)3 sample volume. *Science (80-. ).* **339**, 561–563 (2013).
26. Davis, H. C. *et al.* Mapping the microscale origins of magnetic resonance image contrast with subcellular diamond. *Nat. Commun.* **9**, 1–9 (2018).
27. Ermakova, A. *et al.* Detection of a Few Metallo-Protein Molecules Using Color Centers in Nanodiamonds. *Nano Lett.* **13**, 3305–3309 (2013).
28. Steinert, S. *et al.* Magnetic spin imaging under ambient conditions with sub-cellular resolution. *Nat. Commun.* **4**, (2013).
29. Glenn, D. R. *et al.* Single-cell magnetic imaging using a quantum diamond microscope. *Nat. Methods* **12**, 736–738 (2015).
30. Morita, A. *et al.* Cell Uptake of Lipid-Coated Diamond. *Part. Part. Syst. Charact.* **1900116**, 1–8 (2019).
31. Tetienne, J. P. *et al.* Spin relaxometry of single nitrogen-vacancy defects in diamond nanocrystals for magnetic noise sensing. *Phys. Rev. B - Condens. Matter Mater. Phys.* **87**, 1–5 (2013).
32. Jamieson, D. J. Oxidative stress responses of the yeast Saccharomyces cerevisiae. *Yeast* **14**, 1511–1527 (1998).
33. Juarez, J. C. *et al.* Superoxide dismutase 1 (SOD1) is essential for H2O 2-mediated oxidation and inactivation of phosphatases in growth factor signaling. *Proc. Natl. Acad. Sci. U. S. A.* **105**, 7147–7152 (2008).

34. Reddi, A. R. & Culotta, V. C. SOD1 integrates signals from oxygen and glucose to repress respiration. *Cell* **152**, 224–235 (2013).

35. Che, M. *et al.* Expanding roles of superoxide dismutases in cell regulation and cancer. *Drug Discov. Today* **21**, 143–149 (2016).

36. Tsang, C. K. *et al*. Superoxide dismutase 1 acts as a nuclear transcription factor to regulate oxidative stress resistance. *Nat. Commun.* **5**, 3446 (2014).

37. Betz, C. & Hall, M. N. Where is mTOR and what is it doing there? *J. Cell Biol.* **203**, 563–574 (2013).

38. Li, H. *et al*. Nutrient regulates TOR1 nuclear localization and association with rDNA promoter. *Nature* **442**, 1058–1061 (2006).

39. Beck, T. & Hall, M. N. The TOR signalling pathway controls nuclear localization of nutrient-regulated transcription factors. *Nature* **402**, 689–692 (1999).

40. Crespo, J. L. & Hall, M. N. Elucidating TOR Signaling and Rapamycin Action: Lessons from Saccharomyces cerevisiae. *Microbiol. Mol. Biol. Rev.* **66**, 579–591 (2002).

41. Smith, J. J. & Aitchison, J. D. Peroxisomes take shape. *Nat. Rev. Mol. Cell Biol.* **14**, 803–817 (2013).

42. Sacksteder, K. A. *et al.* PEX19 binds multiple peroxisomal membrane proteins, is predominantly cytoplasmic, and is required for peroxisome membrane synthesis. *J. Cell Biol.* **148**, 931–944 (2000).

43. Jamieson, D. J. Saccharomyces cerevisiae has distinct adaptive responses to both hydrogen peroxide and menadione. *J. Bacteriol.* **174**, 6678–6681 (1992).

44. Flattery-O'Brien, J., Collinson, L. P. & Dawes, I. W. Saccharomyces cerevisiae has an inducible response to menadione which differs from that to hydrogen peroxide. *J. Gen. Microbiol.* **139**, 501–507 (1993).

45. Martínez, F.P. *et al.* Comparing nanodiamond based relaxometry under biologically relevant conditions and determining free radicals produced in chemical reactions. *(submitted)*, (2020).

46. Semchyshyn, H. M. & Valishkevych, B. V. Hormetic effect of H2O2 in Saccharomyces cerevisiae: Involvement of TOR and glutathione reductase. *Dose-Response* **14**, 1–12 (2016).

47. Götte, K. *et al.* Pex19p, a Farnesylated Protein Essential for Peroxisome Biogenesis. *Mol. Cell. Biol.* **18**, 616–628 (1998).

48. Pascual-Ahuir, A., Manzanares-Estreder, S. & Proft, M. Pro- and Antioxidant Functions of the Peroxisome-Mitochondria Connection and Its Impact on Aging and Disease. *Oxid. Med. Cell. Longev.* **2017**, 1–17 (2017).

49. Herker, E. *et al.* Chronological aging leads to apoptosis in yeast. *J. Cell Biol.* **164**, 501–507 (2004).

50. Reverter-Branchat, G. *et al*. Oxidative damage to specific proteins in replicative and chronological-aged Saccharomyces cerevisiae. Common targets and prevention by calorie restriction. *J. Biol. Chem.* **279**, 31983–31989 (2004).

51. Fabrizio, P. & Longo, V. D. Chronological aging-induced apoptosis in yeast. *Biochim. Biophys. Acta - Mol. Cell Res.* **1783**, 1280–1285 (2008).

52. Harman, D. Aging: a theory based on free radical and radiation chemistry. *J. Gerontol.* **11**, 298–300 (1956).

53. Gladyshev, V. N. The free radical theory of aging is dead. Long live the damage theory. *Antioxidants Redox Signal.* **20**, 727–731 (2014).

54. Mesquita, A. *et al.* Caloric restriction or catalase inactivation extends yeast chronological lifespan by inducing $H_2O_2$ and superoxide dismutase activity. *Proc. Natl. Acad. Sci. U. S. A.* **107**, 15123–15128 (2010).

55. Marchler, G. *et al*. Saccharomyces cerevisiae UAS element controlled by protein kinase A activates transcription in response to a variety of stress conditions. *EMBO J.* **12**, 1997–2003 (1993).

56. Hancock, R. D., Galpin, J. R. & Viola, R. Biosynthesis of L-ascorbic acid (vitamin C) by Saccharomyces cerevisiae. *FEMS Microbiol. Lett.* **186**, 245–250 (2000).

57. Özcan, S. & Johnston, M. Function and Regulation of Yeast Hexose Transporters. *Microbiol. Mol. Biol. Rev.* **63**, 554–569 (1999).

58. Saffi, J. *et al*. Antioxidant activity of L-ascorbic acid in wild-type and superoxide dismutase deficient strains of Saccharomyces cerevisiae. *Redox Rep.* **11**, 179–184 (2006).

59. Shenderova, O. A. *et al.* Review Article: Synthesis, properties, and applications of fluorescent diamond particles. *J. Vac. Sci. Technol. B* **37**, 030802 (2019).

60. Ong, S. Y. *et al*. Shape and crystallographic orientation of nanodiamonds for quantum sensing. *Phys. Chem. Chem. Phys.* **19**, 10748–10752 (2017).

61. Hemelaar, S. R. *et al.* Nanodiamonds as multi-purpose labels for microscopy. *Sci. Rep.* 1–9 (2017). doi:10.1038/s41598-017-00797-2

62. Brachmann, C. B. *et al.* Designer deletion strains derived from Saccharomyces cerevisiae S288C: A useful set of strains and plasmids for PCR-mediated gene disruption and other applications. *Yeast* **14**, 115–132 (1998).

63. Hemelaar, S. R. *et al.* Generally Applicable Transformation Protocols for Fluorescent Nanodiamond Internalization into Cells. *Sci. Rep.* **7**, 5862 (2017).

64. Karas, B. J. *et al.* Transferring whole genomes from bacteria to yeast spheroplasts using entire bacterial cells to reduce DNA shearing. *Nat. Protoc.* **9**, 743–750 (2014).

65. Giaever, G. *et al.* Functional profiling of the Saccharomyces cerevisiae genome. *Nature* **418**, 1–

5 (2002).

66. Morita, A. *et al*. The Fate of Lipid-Coated and Uncoated Fluorescent Nanodiamonds during Cell Division in Yeast. *Nanomaterials* **10**, 1–15 (2020).
67. Loretz, M., Rosskopf, T. & Degen, C. L. Radio-frequency magnetometry using a single electron spin. *Phys. Rev. Lett.* **110**, 1–5 (2013).
68. Bar-Gill, N. *et al.* Suppression of spin-bath dynamics for improved coherence of multi-spin-qubit systems. *Nat. Commun.* **3**, (2012).
69. Rondin, L. *et al.* Stray-field imaging of magnetic vortices with a single diamond spin. *Nat. Commun.* **4**, 1–5 (2013).
70. Rioux, J. A., Levesque, I. R. & Rutt, B. K. Biexponential longitudinal relaxation in white matter: Characterization and impact on T1 mapping with IR-FSE and MP2RAGE. *Magn. Reson. Med.* **75**, 2265–2277 (2016).
71. Canetta, E., Walker, G. M. & Adya, A. K. Nanoscopic Morphological Changes in Yeast Cell Surfaces Caused by Oxidative Stress : An Atomic Force Microscopic Study. *J. Microbiol. Biotechnol* **19**, 547–555 (2009).
72. Teparić, R. *et al*. Increased mortality of Saccharomyces cerevisiae cell wall protein mutants. *Microbiology* **150**, 3145–3150 (2004).
73. Pérez-Gallardo, R. V. *et al*. Reactive oxygen species production induced by ethanol in Saccharomyces cerevisiae increases because of a dysfunctional mitochondrial iron – sulfur cluster assembly system. *FEMS Yeast Res.* **13**, 804–819 (2013).

**Supplementary informations**

**Testing stress conditions with conventional methods**

To validate and compare our magnetometry data we have conducted measurements with conventional methods under the same conditions.

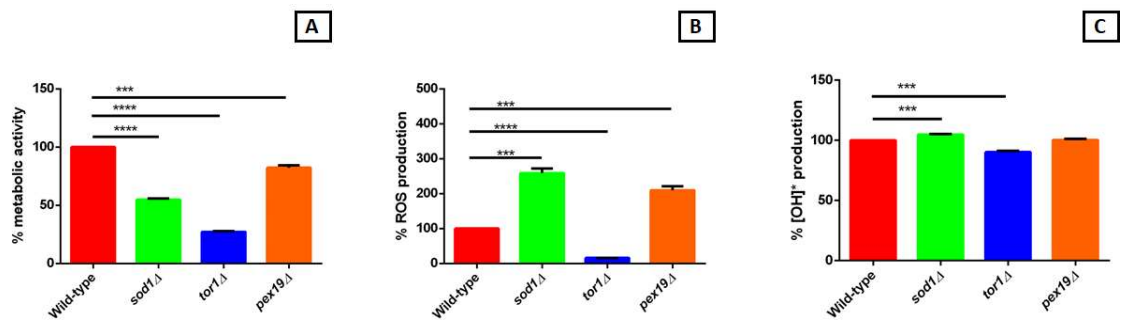

**Supplementary Fig. 1** Characterization of 4 different strains in young stage. MTT result **(A)** shows that all mutant strains have lower metabolic activity compared to the wild-type strain. Those strains then were tested with $H_2$DCFDA to show total ROS production levels **(B)**. As consequence of lacking SOD1 gene, *sod1Δ* produces more ROS than the wild type strain while lacking the TOR1 gene (*tor1Δ*) decreases ROS production. Hydroxyl radical ([OH*]) levels in all strains **(C)** were tested by using an HPF probe. As expected, *sod1Δ* has a higher radical load than the other strains. Wild type yeast was used as a control and set to 100%. The data were analyzed by using a one-way ANOVA followed by a Tukey post hoc test. The measurements were repeated 5 times. The significance level was set to 0.05. P values ≤ 0.001 was indicated with *** and P values ≤0.0001 was indicated with ****.

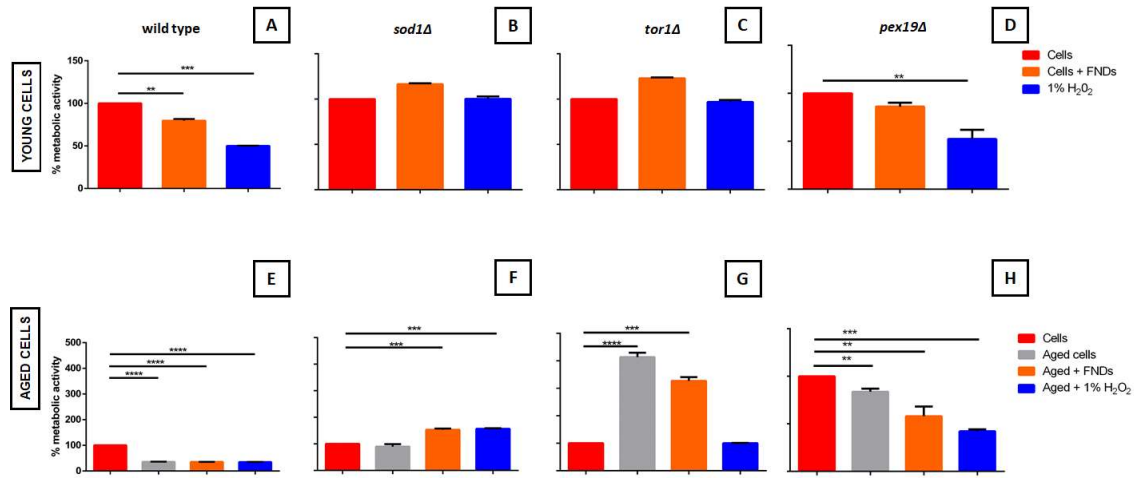

**Supplementary Fig.2** Metabolic activity profile of 4 different strains. All strains were tested at 2 stages (young and aged). The values for cells without treatment (represented in red) were calculated from each strain in their young stage. These were used as control and set to 100%. In the young stage, exposure to FNDs (orange) only has mild effect **(A).** In *sod1Δ* **(B)**, *tor1Δ* **(C)**, and *pex19Δ* **(D)**, the presence of FNDs has no significant effect on metabolic activity levels of the cells. When the cells were triggered with 1% $H_2O_2$ (blue), all strains are showing decreasing metabolic activity. When entering the aged stage, metabolic activity in wild type **(E)**, *sod1Δ* **(F)**, *pex19Δ* **(H)** decrease while *tor1Δ* **(G)** is increasing their metabolic activity compared to young stage. Data were analyzed by using a one-way ANOVA followed by a Tukey post hoc test. The group with cells only was used as a comparison. The measurements were repeated 5 times. Significance level was set to 0.05. (** P values ≤ 0.01 ; *** P values ≤0.001; and  **** P values ≤0.0001).

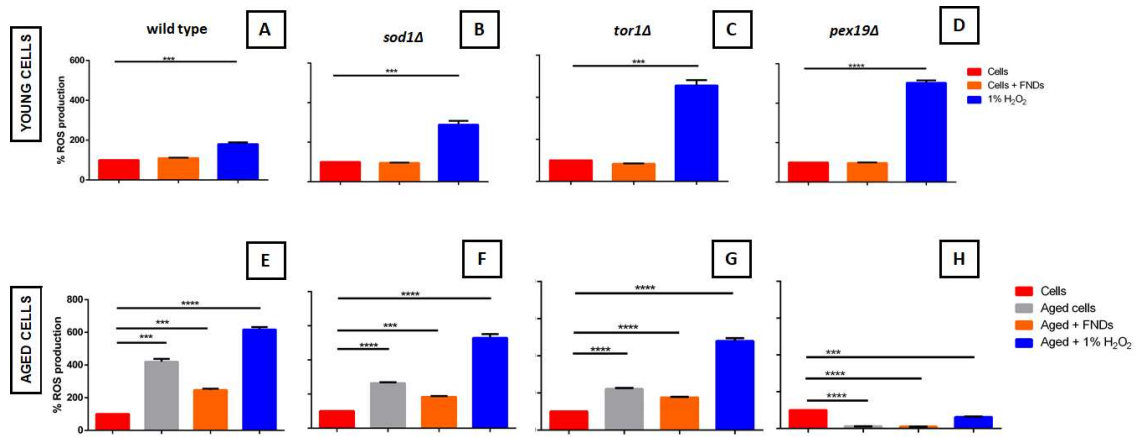

**Supplementary Fig.3** ROS production of 4 different strains. All strains were tested in young and aged conditions. The groups with cells only at young stage from each strain were used as control and set to 100% (red). In the young stage, FNDs (orange) has no effect on the ROS production of any of the strains (wild type **(A)**, *sod1Δ* **(B)**, *tor1Δ* **(C)**, and *pex19Δ* **(D)**). Triggering cells with 1% $H_2O_2$ (blue) increases the ROS production inside the cells. When the cells aged, the ROS values increase in wild type **(E)**, *sod1Δ***(F)**, and *tor1Δ***(G)** while in *pex19Δ***(H)** ROS values decrease. The data were analyzed by using a one way ANOVA followed by a Tukey post hoc test. The measurements were repeated 5 times. Significance level was set at 0.05. (*** P values ≤0.001 and  **** P values ≤0.0001).

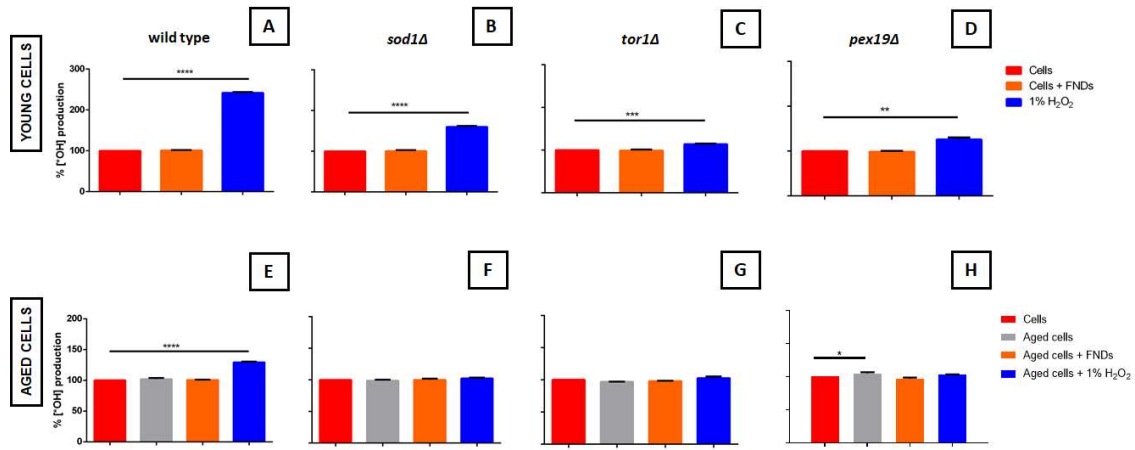

**Supplementary Fig.4** [*OH] level of 4 different strains. All strains were tested in young and aged conditions. The groups with cells only at young stage from each strain were used as control and set to 100% (red). In the young stage FNDs (orange) have no significant effect on [*OH] production on any strain (wild type **(A)**, *sod1Δ* **(B)**, *tor1Δ* **(C)**, and *pex19Δ* **(D)**). Triggering cells with 1% $H_2O_2$ (blue) increases [*OH] production inside the cells. When the cells aged (grey), only *pex19Δ* **(H)** shows increased [*OH] levels while wild type **(E)**, *sod1Δ***(F)**, and *tor1Δ***(G)** remain the same. When aged cells were triggered with 1% $H_2O_2$ (blue), only the wild type has significant elevation of [*OH]. The data were analyzed by using a one way ANOVA followed by a Tukey post hoc test. The measurements were repeated 5 times. Significance level was set at 0.05. (* P values ≤0.05; ** P values ≤0.01; *** P values ≤0.001 and **** P values ≤0.0001).

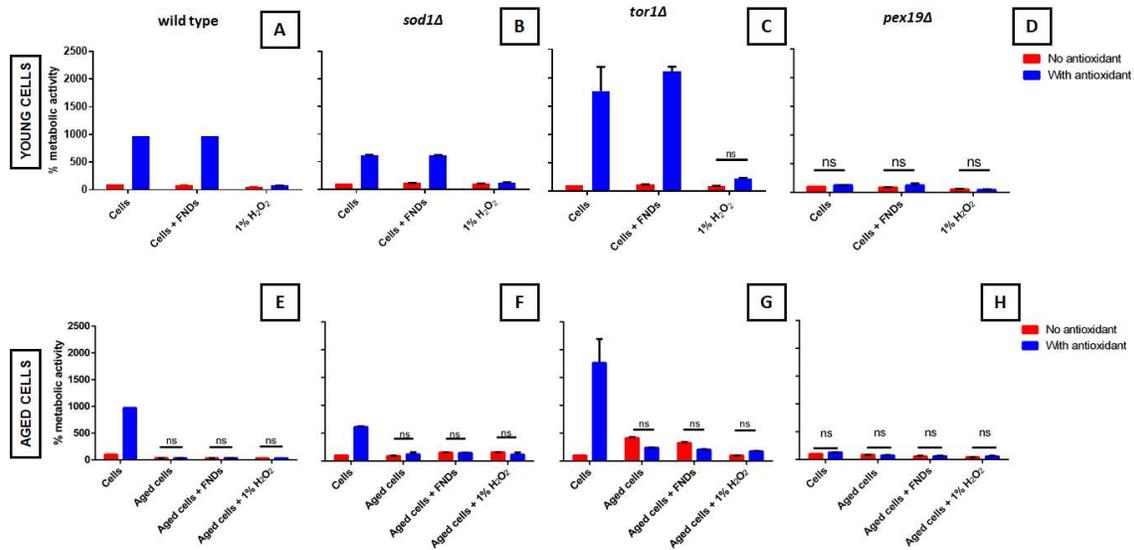

**Supplementary Fig.5** Effect of antioxidant supplement in metabolic activity profile. Yeasts were induced with antioxidant to prevent oxidative stress during the aging process. In the young stage, adding antioxidant (blue) in wild type **(A)**, *sod1Δ* **(B)**, and *tor1Δ* **(C)** can increase the metabolic activity. On the other hand, in *pex19Δ* **(D)**, adding antioxidant almost has no effect on improving metabolic activity. When entering the aging stage, antioxidant only has an effect (increasing metabolic activity) on cell wild type cells **(E)**, *sod1Δ* **(F)**, and *tor1Δ* **(G)**. However, the antioxidant had no significant effect on pex19Δ **(H)**. The data were analyzed by using two-way ANOVA followed by a Sidak post hoc test. The significance level was set at 0.05. The measurements were repeated 5 times. (ns means not significant).

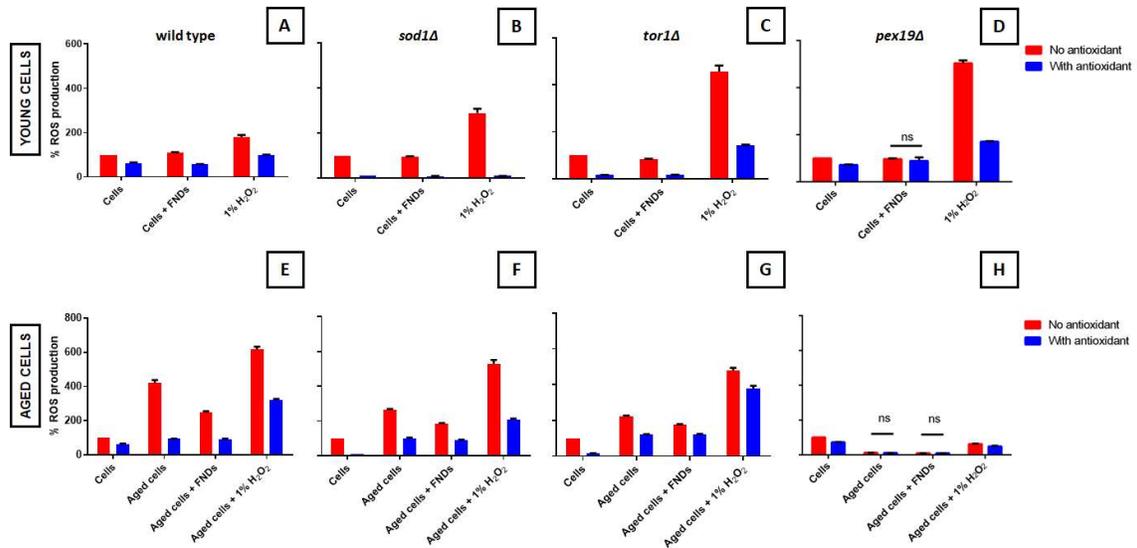

**Supplementary Fig.6** Effect of antioxidant supplement on ROS production. All strains were induced with antioxidant. In the young stage, adding antioxidant (blue) in wild type **(A)**, *sod1Δ* **(B)**, *tor1Δ* **(C)**, and *pex19Δ* **(D)** can decrease ROS production. When entering the aging stage, antioxidant has a significant effect on reducing ROS production in wild type **(E)**, *sod1Δ* **(F)**, and *tor1Δ* **(G)**. The effect of antioxidant in aged *pex19Δ* **(H)** is not significant. The data were analyzed by using a two-way ANOVA followed by a Sidak post hoc test. The significance level was set at 0.05. The measurements were repeated 5 times. (ns means not significant).

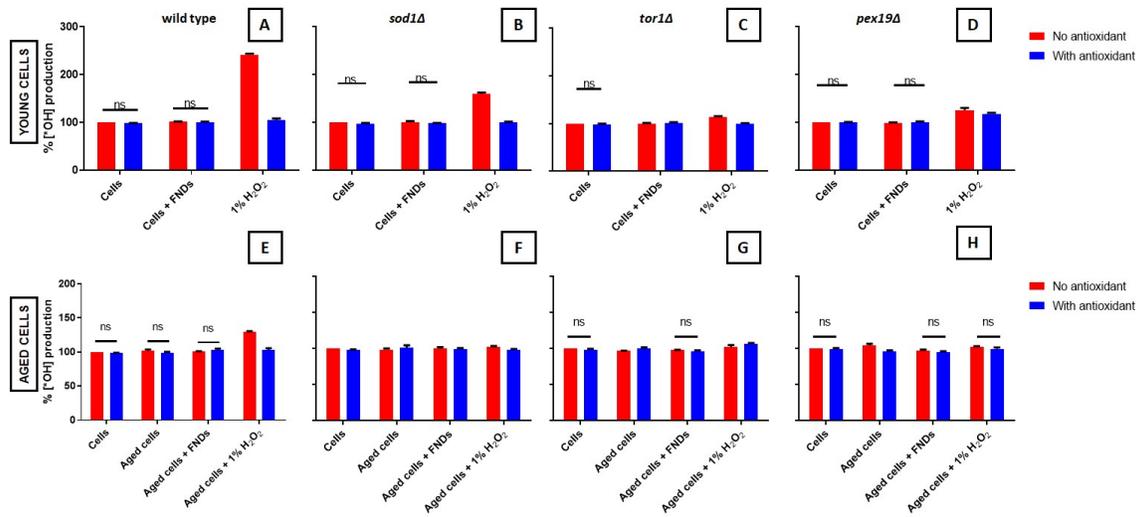

**Supplementary Fig.7** Effect of antioxidant supplement in [*OH] production. All strains were induced with antioxidant. In the young stage, adding antioxidant (blue) in wild type **(A)**, *sod1Δ* **(B)**, *tor1Δ* **(C)**, and *pex19Δ* **(D)** has no effect on the [*OH] concentrations. Effect of antioxidant on reducing [*OH] can be seen in cells with 1% $H_2O_2$ treatment for all strains. In the aging stage, antioxidant has almost no effect on reducing [*OH] production in the wild type **(E)**, *tor1Δ* **(G)**, and *pex19Δ* **(H)** while in *sod1Δ* **(G)**, adding antioxidant slightly reduces [*OH]. The data were analyzed by using two-way ANOVA followed by a Sidak post hoc test. The significance level was set at 0.05. The measurements were repeated 5 times. (ns means not significant).

**The effect of chemicals on bare diamonds**

To exclude that the used chemicals induce a change in T1 by themselves we tested FNDs with the chemicals alone without the involvement of cells.

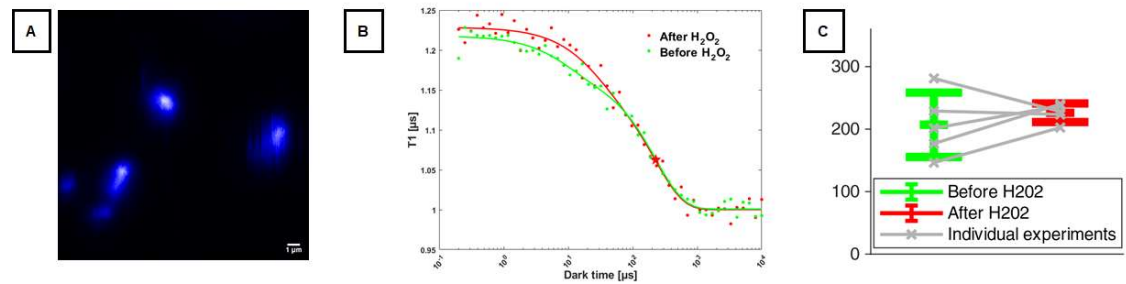

**Supplementary Fig.8 (A)** shows a confocal image of bare FNDs taken from the home-build magnetometry setup. The scale bar is 1 μm. The initial T1 values of FNDs were recorded then 1% $H_2O_2$ was added to the particles. **(B)** shows representative T1 curves. T1 values for both groups were compared in **(C)**. There is no significant difference between the measurement before and after adding 1% $H_2O_2$. Data were analyzed using a paired T-test and significance level was set at 0.05.

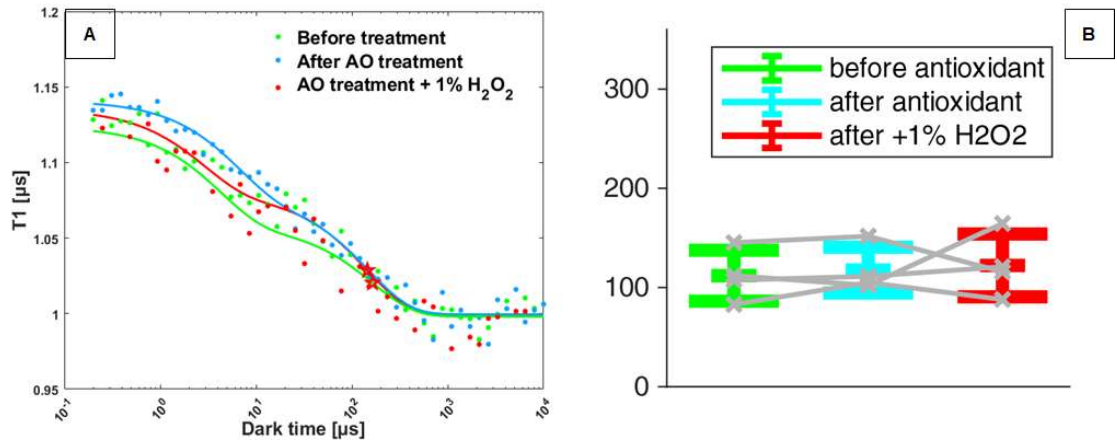

**Supplementary Fig.9** Effect of anti-oxidant on bare FNDs without cells. **(A)** Representative T1 curves were obtained from bare FNDs with anti-oxidant followed by 1% $H_2O_2$. FNDs were measured before (in green) adding any chemicals and used as a control group. The T1 was measured after adding 0.025 mM L-ascorbic acid (blue) followed by adding 1% $H_2O_2$ (red). **(B)** T1 values were collected and analyzed by using a one-way ANOVA followed by a Tukey post hoc test. According to the statistical test, the differences between the groups were not significant (P value 0.53). Data were analyzed by a paired T-test against previous group and P value < 0.05 was considered has significant difference. AO means anti-oxidant.

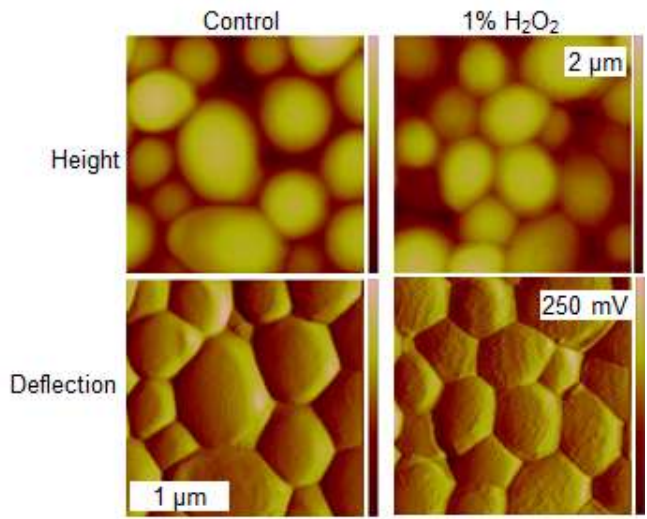

**Supplementary Fig.10** Morphological changes during oxidative stress in yeasts. Yeast cells were stimulated by 1% $H_2O_2$ then we observed the morphological changes using AFM. Yeasts without any stress inducer have a smooth surface while stressed yeasts have a rougher (wrinkly) surface.

Comparison between different methods

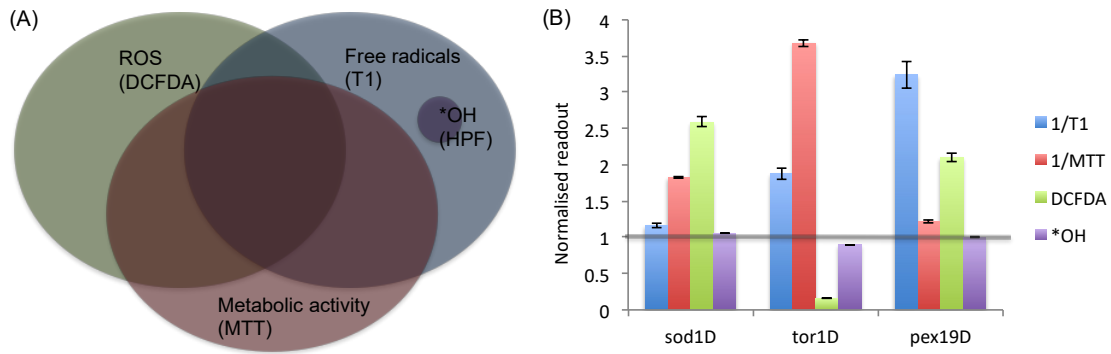

**Supplementary Fig.11:** Comparison between different methods. (A) shows what the respective techniques in this work are measuring exactly and how they are related. The colored circles stand for different groups of molecules that are detected. They do have a certain overlap but there is currently no technique, which is able to measure the exact same signal as we do. (B) compares the data in Fig 2 (B) of the main manuscript with the existing techniques. For a better comparison we have normalized all the data to the average value for the wild type strain with each technique (indicated by the grey line). Further, we are showing 1/T1 and 1/MTT. As a result, in this graph the depicted values are all proportional to the concentration of measured molecules.

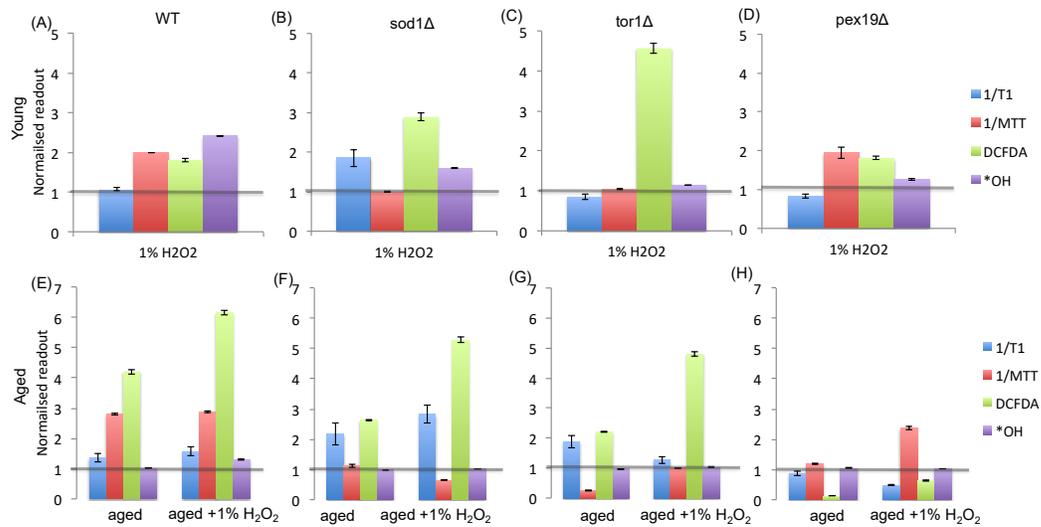

**Supplementary Fig.12:** Comparison between different methods. Here we compare the data in Fig 3 of the main manuscript with the existing techniques. Again all the data are normalized all but this time to the initial measurements with each technique (indicated by the grey line). Also here, we are showing 1/T1 and 1/MTT. As a result, in this graph the depicted values are all proportional to the concentration of measured molecules.

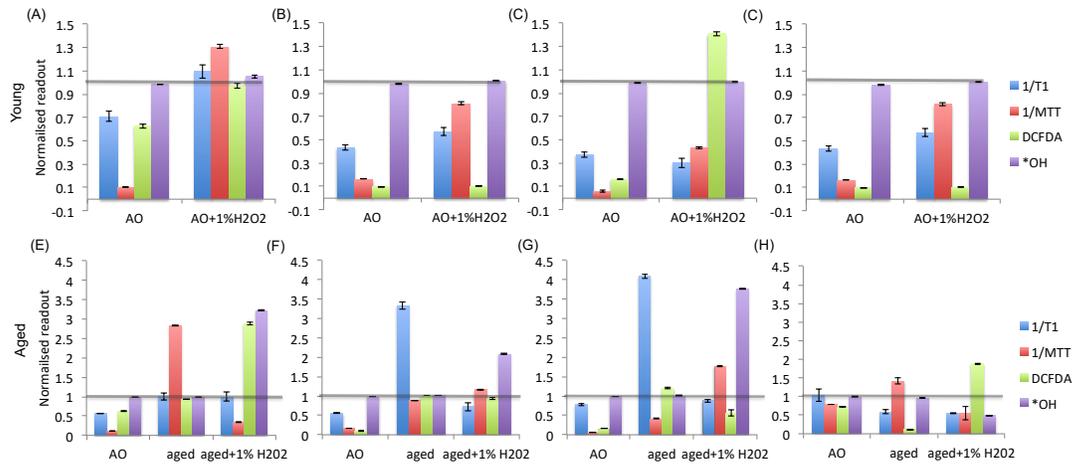

**Supplementary Fig.13:** Comparison between different methods. Here we compare the data in Fig 4 of the main manuscript with the existing techniques. Again all the data are normalized all but this time to the initial measurements with each technique (indicated by the grey line). Also here, we are showing 1/T1 and 1/MTT. As a result, in this graph the depicted values are all proportional to the concentration of measured molecules.